\documentclass[twocolumn,prb,amsmath,amssymb]{revtex4}
\usepackage{graphicx}
\usepackage{booktabs}
\usepackage{multirow}
\usepackage{amssymb}
\usepackage{graphicx}
\usepackage{amsmath}
\usepackage{indentfirst}
\usepackage{pstricks}
\usepackage{verbatim}
\usepackage{bm}
\usepackage{color}

\begin{document}
\title
{Ink-Jet Printed Graphene Electronics}
\author{F. Torrisi, T. Hasan, W. Wu, Z. Sun, A. Lombardo, T. Kulmala, G. W. Hshieh, S. J. Jung, F. Bonaccorso, P. J. Paul, D. P. Chu, A. C. Ferrari}\email{acf26@eng.cam.ac.uk}

\affiliation{Department of Engineering, University of Cambridge, Cambridge CB3 0FA, UK}

\begin{abstract}
We demonstrate ink-jet printing as a viable method for large area fabrication of graphene devices. We produce a graphene-based ink by liquid phase exfoliation of graphite in N-Methylpyrrolidone. We use it to print thin-film transistors, with mobilities up to$\sim$95cm$^{2}$V$^{-1}$s$^{-1}$, as well as transparent and conductive patterns, with$\sim$80$\%$ transmittance and$\sim$30k$\Omega$/$\Box$ sheet resistance. This paves the way to all-printed, flexible and transparent graphene devices on arbitrary substrates.
\end{abstract}
\maketitle
\section{\label{In}Introduction}
Flexible electronics is a rapidly expanding research area\cite{Cao2008}. Applications include touch screens\cite{Zhou2006}, electronic paper (e-paper)\cite{Ota1973,Gelinck2004}, sensors\cite{Sekitani2009}, radio frequency tags\cite{Myny2009}, photovoltaic cells\cite{Granqvist2007,Yoon2008}, and electronic textiles\cite{Schmied2009}. To date, it mainly relies on two fabrication strategies: one in which substrates bearing thousands of Field-effect Transistors (FETs) are bonded to plastic by transfer printing or pick-and place methods\cite{kim2008}; another in which FETs are prepared directly on the target substrate by several coating, curing and lithographic steps\cite{Singh2006,Cao2008}. Rubber stamping\cite{Rogers2001}, embossing\cite{Forrest2004} and ink-jet printing\cite{Bao1999,Sirringhaus2000} reduce the number of such fabrication steps.

Ink-jet printing is one of the most promising techniques for large area fabrication of flexible plastic electronics\cite{Sirringhaus2000}. A range of components can be printed, such as transistors\cite{Forrest2004,Sirringhaus2000,Sun2006,McAlpine2005,Singh2010}, photovoltaic devices\cite{Peumans2003}, organic light emitting diodes (OLEDs)\cite{Forrest2004,Servati2005,Singh2010}, and displays\cite{Forrest2004}. Ink-jet printing is versatile\cite{Singh2010}, involves a limited number of process steps\cite{deGans2004}, is amenable for mass production, and can deposit controlled amounts of material\cite{deGans2004}. Drop on demand\cite{deGans2004,Dong2006} ink-jet printing has progressed from printing text and graphics\cite{deGans2004}, to a tool for rapid manufacturing\cite{van Osch2008}, being now an established technique to print Thin Film Transistor (TFT) based on organic conducting and semiconducting inks\cite{Sirringhaus2000,Sekitani2009,Yoo2010}. However, their mobilities, $\mu$$<$0.5cm$^{2}$V$^{-1}$s$^{-1}$,\cite{Sekitani2009,Singh2010} are still much lower than standard silicon technology. Several approaches aim to improve this, such as the use of polysilicon\cite{Shimoda2006}, zinc oxide nanoparticles\cite{Noh2007} and carbon nanotubes (CNTs)\cite{Beecher2007,Hsieh2009,Takenobu2009,Okimoto2009,Okimoto2010,Ha2010}. Metal nanoparticle inks are not stable in ordinary solvents, such as Deionized (DI) Water, Acetone, Isopropyl Alcohol, N-Methylpyrrolidone (NMP), Tetrahydrofuran\cite{Singh2010,Luechinger2008}. They need to be chemically modified in order to be dispersed\cite{Singh2010}, using stabilizers, which usually degrade in a couple of years\cite{Singh2010,Luechinger2008}. Metal nanoparticles also tend to oxidize after printing\cite{Singh2010,Luechinger2008}. Ink-jet printed CNT-TFTs have been reported with $\mu$ up to 50cm$^{2}$ V$^{-1}$s$^{-1}$ and a ON/OFF ratio$\sim$10$^{3}$.\cite{Ha2010}

Graphene is the two-dimensional (2d) building block for $sp^{2}$ carbon allotropes of every other dimensionality. It can be stacked into 3d graphite, rolled into 1d nanotubes, or wrapped into 0d fullerenes\cite{Geim_rise2007}. It is at the centre of an ever expanding research area\cite{Geim_rise2007,Novoselov2004,Charlier2008,Bonaccorso2010}. Near-ballistic transport and high mobility make it an ideal material for nanoelectronics, especially for high frequency applications\cite{Lin2010}. Furthermore, its optical and mechanical properties are ideal for micro and nanomechanical systems, thin-film transistors, transparent and conductive composites and electrodes, and photonics\cite{Geim_rise2007,Bonaccorso2010,sunacsnano10}. Graphene was isolated by micromechanical exfoliation of graphite\cite{NovoselovPNAS2005}. This technique is still the best in terms of purity, defects, mobility and optoelectronics properties. However, large scale production approaches are needed for widespread application. These encompass growth by chemical vapor deposition (CVD)\cite{Karu1966,Obraztsov2007,Kim2009,Reina2009,Li2009,Bae2010}, segregation by heat treatment of silicon carbide\cite{Berger2006,Acheson1896,Badami1962,Emtsev2009} and metal substrates\cite{Oshima1997,Gamo1997,Rosei1983,Sutter2008}, liquid phase exfoliation (LPE)\cite{Hernandez2008,Lotya2009,Valles2008,Hasan2010}. Amongst these, LPE is ideally suited to produce printable inks.

Graphite can be exfoliated by chemical wet dispersion followed by ultrasonication, both in aqueous\cite{Lotya2009,Hasan2010} and non-aqueous solvents\cite{Hernandez2008,Hasan2010}. Dispersions can be achieved by mild sonication in water with Sodium Deoxycholate, followed by sedimentation based-ultracentrifugation\cite{Hasan2010,Marago2010}. Bile salt surfactants also allow the isolation of flakes with controlled thickness, when combined with density gradient ultracentrifugation (DGU)\cite{Greennlett2009}. Exfoliation of graphite intercalated compounds\cite{Valles2008} and expandable graphite\cite{LiScience2008} was also reported.

LPE was first achieved through sonication of graphite oxide\cite{Stankovich2006}, following the Hummers method\cite{Hummers1958}. The oxidation of graphite in the presence of acids and oxidants\cite{Brodie1860,Staudenmaier1898} disrupts the sp$^{2}$-network and introduces hydroxyl or epoxide groups\cite{Mattevi2009,Cai2008}, with carboxylic or carbonyl groups attached to the edge\cite{Mattevi2009,Cai2008}. These make graphene oxide (GO) sheets readily dispersible in water\cite{Eda2010,Stankovich2006} and several other solvents\cite{GO_organic_08}. Although large GO flakes can be produced, these are intrinsically defective\cite{Stankovich2006,he_GO_model_98}, and electrically insulating\cite{Stankovich2006,Mattevi2009}. Despite several attempts\cite{Stankovich2006,Mattevi2009}, reduced GO (RGO) does not fully regain the pristine graphene electrical conductivity\cite{Eda2008,Mattevi2009}. It is thus important to distinguish between dispersion processed graphene flakes\cite{Hernandez2008,Lotya2009,Valles2008,Hasan2010}, retaining the electronic properties of graphene, and insulating GO dispersions\cite{Stankovich2006,Eda2008}. Several groups reported GO-based inks\cite{Dua2010,Wang2009,Luechinger2008}. Ref. \onlinecite{Dua2010} ink-jet printed RGO films for sensors applications, while Ref. \onlinecite{Luechinger2008} produced RGO-stabilized Cu nanoparticles as low temperature metal colloids, to replace standard metal nanoparticle inks, that require high temperature sintering postprocessing\cite{Park2007}. Mobilities up to 90cm$^{2}$V$^{-1}$s$^{-1}$ have been achieved for highly reduced GO films by ink-jet printing\cite{Wang2009}, with an ON/OFF ratio up to 10.\cite{Wang2009}

Here we produce a graphene-based ink and demonstrate its viability for printed electronics.
\section{Results and discussion}
\section{Ink requirements}
A key property of inks viable for printing is their ability to generate droplets\cite{Reis2000,Jang2009}. Ink viscosity, $\eta$ [mPa s], surface tension, $\gamma$ [mJ m$^{-2}$], density, $\rho$ [g cm$^{-3}$], and nozzle diameter, $a$ [$\mu$m], influence the spreading of the liquid drops\cite{Fromm1984,Reis2000,Jang2009}. These parameters can be arranged into dimensionless figures of merit (FOM), such as the Reynolds (Re)\cite{Fromm1984,Reis2000,Jang2009}, Weber
(We)\cite{Fromm1984,Reis2000,Jang2009}, and Ohnesorge (Oh)\cite{Fromm1984,Reis2000,Jang2009} numbers: Re=$\frac{\upsilon \rho a}{\eta}$; We=$\frac{\upsilon^{2}\rho a}{\gamma}$, Oh=$\frac{\sqrt{We}}{Re}=\frac{\eta}{\sqrt{\gamma\rho a}}$, where $\upsilon$[m/s] is the drop velocity.

Refs. \onlinecite{Fromm1984,Reis2000,Jang2009} suggested to use Z=1/Oh as the appropriate FOM to characterize drop formation, 1$<$Z$<$14 being required to get stable drop generation\cite{Reis2000,Jang2009}. For Z$<$1 the high viscosity prevents drop ejection\cite{Reis2000,Jang2009}, whereas at Z$>$14 the primary drop is accompanied by a number of satellite droplets\cite{Reis2000,Jang2009}. Moreover, when inks contain dispersed molecules or nano-particles, the latter should be smaller than the nozzle diameter, to prevent clogging\cite{van Osch2008,deGans2004}. Refs. \onlinecite{van Osch2008,Microfab1999} suggested that the size of such molecules or particles should be at least 1/50 of the nozzle diameter, in order to exclude any printing instability, such as clustering of the particles at the nozzle edge, which may deviate the drop trajectory, or result in agglomerates that will eventually block the nozzle.

The ejected drop behavior on the substrate can be efficiently described by fluid dynamics. When a small liquid droplet is in contact with a flat surface, partial wetting results in a finite angle between the liquid and the substrate\cite{deGennes1985}, known as contact angle, $\theta_{C}$\cite{Shafrin1960,deGennes1985,Israelachvili1991}. The lower drop size limit is given by\cite{Reis2000,Jang2009} $s[\mu m]=a\sqrt{\frac{We+12}{3(1-cos\theta_{C})+4We/Re^{1/2}}}$. Thus, e.g., for a typical $a$=50$\mu$m, $We$=20, $Re$=58 and $\theta_{C}$$\sim$45$^\circ$, we get $s$$\sim$85-90$\mu$m. The distance from the substrate must be optimized to guarantee both homogeneous printing and the highest resolution, barring any unusual jetting conditions, such as perturbations from the surrounding environment and diversion of the drop trajectory\cite{Reis2000,Derby2003,Singh2010}. Furthermore, a substrate very close to the nozzle causes secondary drops to scatter off during the impact of the primary drop\cite{Singh2010,Park2010}, due to the initial drop jetting pressure, thus affecting the homogeneity of the final printed features\cite{Park2010}. The final assembly of printed nano-particle inks depends on the substrate Surface Energy (SE)\cite{deGans2004,van Osch2008}, as well as the ink viscosity and surface tension\cite{deGans2004}.

When a drop of an ink containing dispersed particles evaporates on a surface it commonly leaves a dense, ring-like, deposit along its perimeter\cite{deGans2004,van Osch2008}. This is the so-called "coffee ring effect"\cite{Deegan1997}, i.e. a distortion of the drops during solvent drying due to the interplay of ink viscosity and solute transport via solvent motion (arising from surface tension interaction between solvent and substrate)\cite{Deegan1997,Singh2010}. This is one of the most important phenomena affecting the homogeneity of ink-jet printed drops\cite{Deegan1997,Singh2010}. In order to prevent this, it is necessary to "freeze" the drops geometry immediately after they form an homogeneous and continuous film on the substrate\cite{Singh2010}.

Here we use an ink-jet printer with a nozzle diameter$\sim$50$\mu$m, thus we need to have flakes less than 1$\mu$m across. By tuning $\eta$, $\gamma$ and $\rho$ we will target a Z within the optimal range. We print on Si/SiO$_{2}$ (to probe the electrical properties of the ink) and borosilicate (Pyrex 7740-Polished Prime Grade) glass substrates (to test the ink as transparent conductor), both with a roughness R$_{z}$$<$15nm. Our aim is to obtain ink-jet printed drops on the substrate, with homogeneous flakes and uniform morphology, i.e. with roughness comparable to the substrate. We obtain this by varying the contact angle and optimizing the substrate wettability.

In order to reduce the coffee ring effect we need both a solvent with boiling point (T$_{c}$ [$^\circ$C]) and heat of vaporization (V$_{c}$ [kJ/mol]) higher than water\cite{Deegan1997,Derby2003,Singh2010}, and a substrate that promotes adhesion\cite{Osthoff2007}. Thus, we use NMP as solvent for two main reasons. First, it has higher boiling point ($\sim$202$^\circ$C)\cite{HandchemLide} and heat of vaporization (54.5kJ/mol)\cite{HandchemLide}, than water ($\sim$100$^\circ$C and $\sim$40kJ/mol). Second, NMP is the best solvent to get high-yield, surfactant-free exfoliation of graphite\cite{Hernandez2008,Hasan2010}. We then test several surface treatments to optimize substrate adhesion. After printing, NMP is removed by thermal annealing at 170$^\circ$C for 5 minutes.
\subsection{Graphene-based printable ink}
We prepare the graphene-based printable ink as follows. Graphite flakes (NGS Naturgraphit) are sonicated (Decon bath, 100W) in NMP for 9 hours. The un-exfoliated flakes are left to settle for 10 mins after sonication. The decanted dispersions are then ultracentrifuged using a TH-641 swinging bucket rotor in a Sorvall WX-100 Ultra-centrifuge at 10,000 rpm ($\sim$15,000g) for an hour and filtered to remove flakes$>$1$\mu$m, that might clog the nozzle. The resulting ink is characterized by Optical Absorption Spectroscopy (OAS), High Resolution Transmission Electron Microscopy (HRTEM), Electron diffraction and Raman spectroscopy.
\begin{figure}
\centerline{\includegraphics[width=80mm]{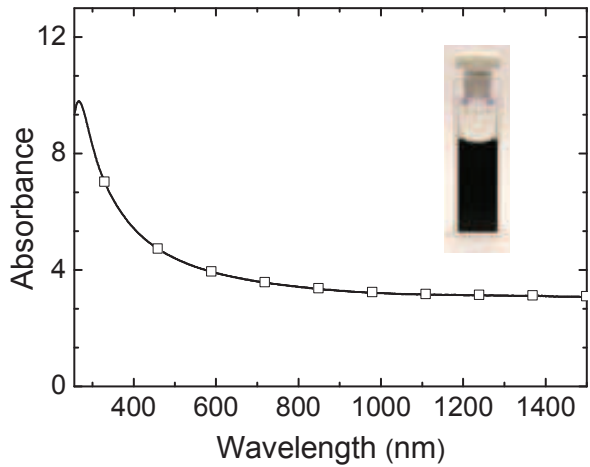}}
\caption{\label{absorption} Absorbance of graphene-ink. The inset is a picture of a vial of ink.}
\end{figure}

A Perkin-Elmer Lambda 950 spectrometer with 1nm resolution is used for OAS measurements. OAS can be used to estimate the concentration of graphene\cite{Hernandez2008,Lotya2009,Marago2010} using the Beer-Lambert Law according to the relation $A=\alpha cl$, where A is the absorbance, l [m] is the light path length, c [g/L] the concentration of dispersed graphitic material and $\alpha$ [L g$^{-1}$ m$^{-1}$] the absorption coefficient. Fig.\ref{absorption} plots an OAS spectrum of our ink diluted to 10\%. The ink is diluted to avoid strong scattering losses at higher concentrations, which could cause deviation of the measured value of A from the Beer-Lambert law. The spectrum in Fig.\ref{absorption} is mostly featureless, as expected due to the linear dispersion of the Dirac electrons\cite{sunacsnano10,makprl08,kravetsprb10,Bonaccorso2010,Nair2008,Casiraghi2007}, the peak in the UV region being a signature of the van Hove singularity in the graphene density of states\cite{kravetsprb10}. From $\alpha\sim$1390Lg$^{-1}$m$^{-1}$ at 660nm, as for Refs. \onlinecite{Lotya2009,Hasan2010}, we estimate c$\sim$0.11$\pm$0.02g/L.
\begin{figure}
\centerline{\includegraphics[width=90mm]{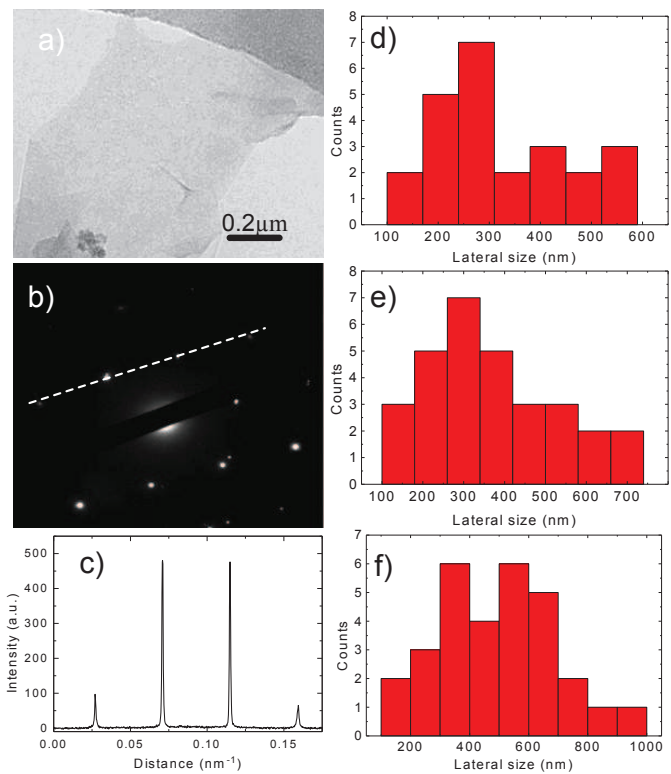}}
\caption{\label{TEM} a,b) HRTEM image and diffraction pattern of a dispersed SLG flake. c) Diffracted intensity along the dashed line in b. Lateral size distribution of d) SLGs, e) BLGs, f) FLGs.}
\end{figure}

We disperse drops of our ink on Holey carbon Transmission electron microscopy (TEM) grids for analysis using a Tecnai T20 high resolution TEM, with an acceleration voltage of 200KV operating in phase contrast mode. Fig.\ref{TEM}a is HRTEM image of a Single Layer Graphene (SLG) flake from the ink, while Fig.\ref{TEM}b is a normal-incidence electron diffraction of the same flake of Fig.\ref{TEM}a. It shows the expected sixfold symmetry\cite{Meyer_nature2007,Meyer_SSC2007,Ferrari2006}. The peaks are labeled with the corresponding Miller-Bravais (hkil) indexes. For Few Layer Gaphene (FLG) flakes with Bernal (AB) stacking, the intensity ratio $I_{1100}/I_{2110}$ is$<$1, while for SLG $I_{1010}/I_{2110}$$>$1\cite{Meyer_nature2007,Ferrari2006}. We use this to distinguish SLG from FLGs\cite{Hernandez2008,Marago2010}. Fig.\ref{TEM}c plots the diffraction intensity measured along the line section through the (1$\overline{2}$10), (0$\overline{1}$10), ($\overline{1}$010), ($\overline{2}$110) axis, reported in Fig.\ref{TEM}b. The inner peaks, (0$\overline{1}$10) and ($\overline{1}$010), are$\sim$1.5 times the outer ones, (1$\overline{2}$10) and ($\overline{2}$110), indicating that the flake is SLG\cite{Meyer_nature2007}. The analysis of the edges also gives a reliable information on the number of layers and can be used to investigate a large number of flakes\cite{Meyer_nature2007}, from zoomed-in high resolution edge images\cite{Hernandez2008,Khan2010}. If SLG folds or several SLGs stack one on the other, selected area diffraction is used to distinguish contentious cases.

These combined analysis show that our ink mostly consists of SLGs, Bi-Layers (BLG) and FLGs, with lateral size$\sim$300-1000nm. We find that$\sim$35\% SLGs are larger than 300nm (Fig.\ref{TEM}d);$\sim$40\% BLGs are larger than 350nm (Fig.\ref{TEM}e);$\sim$55\% FLGs are larger than 450nm (Fig.\ref{TEM}f). In particular, we have$\sim$33\% SLG with c$\sim$0.11g/L. Previous works on LPE of graphite in NMP reported up to$\sim$28\% SLG for c$\sim$0.18g/L\cite{Hasan2010} and $\sim$21\% for c$\sim$1.8g/L\cite{Khan2010}. Ref. \onlinecite{Valles2008} also reported exfoliation of intercalated graphite in NMP, with$\sim$20\% SLGs for c$\sim$0.01g/L. Thus, our ink has higher SLG yield with respect to previous works, but lower c than Ref.\onlinecite{Khan2010}. This higher c was achieved by long time (up to 460h) ultrasonication\cite{Khan2010}. However Ref. \onlinecite{Khan2010} reported defects and reduction of size as a result. Our combination of low-power sonication ($<$25W) and ultracentrifugation is ideal for high-yield of defect-free SLGs.

Stable dispersions require the Gibbs free energy of mixing, $\Delta G_{mix}$, to be zero or negative\cite{hansenbook}, where $\Delta G_{mix}=\Delta H_{mix}-K\Delta S_{mix}$, K being the temperature, $\Delta H_{mix}$ the enthalpy of mixing and and $\Delta S_{mix}$ the entropy change in the mixing process\cite{Hernandez2008,hansenbook}. For graphene and nanotubes, $\Delta S_{mix}$ is small\cite{Hernandez2008,Bergin2008}. Therefore, for dispersion and stabilization of graphene in solvents, $\Delta H_{mix}$ needs to be very small. This can be achieved by choosing a solvent whose surface energy is very close to that of graphene\cite{Hernandez2008}. The surface energy of NMP satisfies this requirement and allows efficient exfoliation of graphite. Graphite can also be efficiently exfoliated in water with the use of bile salt surfactants. Ref. \onlinecite{Lotya2010} reported$\sim$20\%SLGs  and c$\sim$0.3g/L SLGs, while Ref. \onlinecite{Marago2010} reported$\sim$60\% SLGs for c$\sim$0.012g/L. The yield can be increased up to$\sim$80\% by density gradient ultracentrifugation\cite{Greennlett2009}. The flake size of LPE graphene in water-surfactant dispersions is on average smaller($\sim$200nm\cite{Lotya2010}, $\sim$30nm\cite{Marago2010}) than thus far reported for NMP($\sim$1$\mu$m\cite{Hasan2010,Hernandez2008}). The viscosity at room temperature of NMP (1.7mPas\cite{HandchemLide}) is higher than water ($\sim$1mPas\cite{HandchemLide}). Larger flakes in a higher viscosity medium (such as NMP) experience higher frictional force\cite{Williams1958,Schuck2000} and sedimentation coefficient\cite{Svedberg1940,Schuck2000}, making it more difficult for them to sediment during ultracentrifugation. This reduces the SLG yield in NMP compared to water.

\begin{figure}
\centerline{\includegraphics[width=90mm]{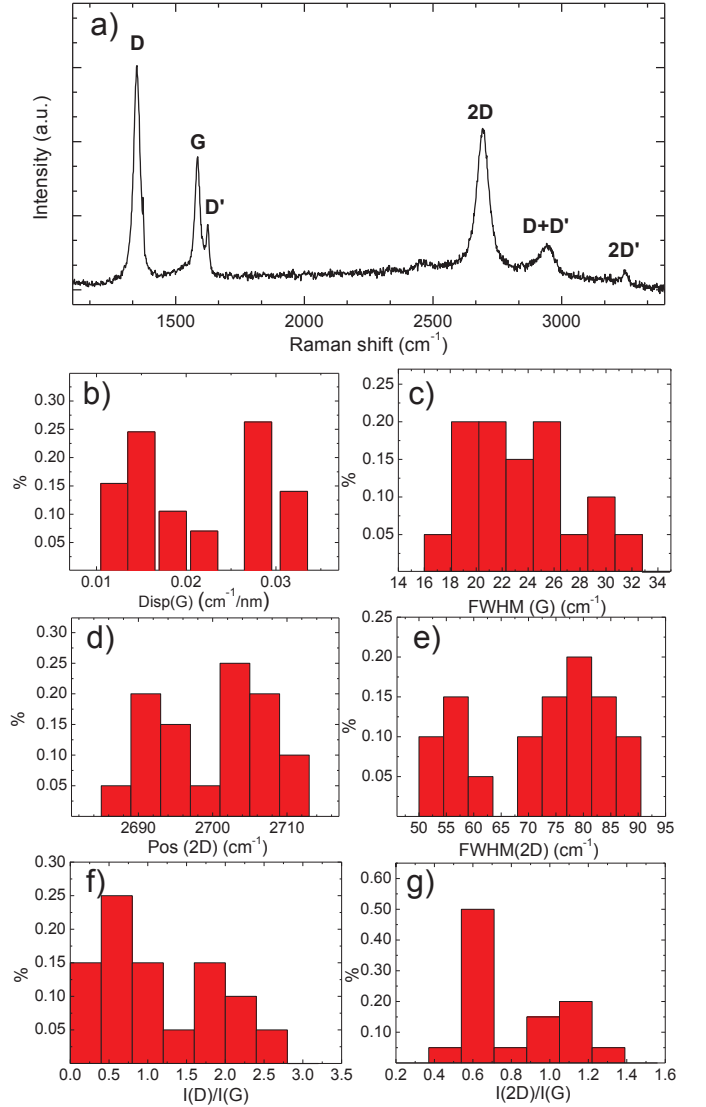}}
\caption{\label{raman} a) Raman spectrum of graphene-ink deposited on Si/SiO$_{2}$. Distribution of b) Disp(G), c) I(D)/I(G), d) FWHM(G), e) Pos(2D), f) FWHM(2D), g) I(2D)/I(G).}
\end{figure}

The centrifuged dispersions are drop-cast onto a Si wafer with 300nm thermally grown SiO$_{2}$ (LDB Technologies ltd.) and annealed at 170$^\circ$C to remove NMP. These samples are then used for Raman measurements, collected with a Renishaw 1000 at 457, 514.5 and 633nm and a 100$\times$ objective, with an incident power$\sim$1mW. Fig.\ref{raman}a plots a typical Raman spectrum of the ink at 514.5nm. Besides the G and 2D peaks, it shows significant D and D' intensities and the combination mode D+D'$\sim$2950cm$^{-1}$. The G peak corresponds to the E$_{2g}$ phonon at the Brillouin zone centre. The D peak is due to the breathing modes of sp$^{2}$ rings and requires a defect for its activation by double resonance (DR)\cite{Ferrari2006,Ferrari2000,Tuinstra1970}. The 2D peak is the second order of the D peak. This is a single band in SLG\cite{Ferrari2006}, whereas it splits in four in BLG, reflecting the evolution of the band structure\cite{Ferrari2006}. The 2D peak is always seen, even when no D peak is present, since no defects are required for the activation of two phonons with the same momentum, one backscattering from the other\cite{Ferrari2006}. DR can also happen intra-valley, {\it i.e.} connecting two points on the same cone around \textbf{K} or \textbf{K'}\cite{Ferrari2000,Tuinstra1970,Piscanec2004}. This gives the D' peak. The 2D' is the second order of the D' peak.
\begin{figure}
\centerline{\includegraphics[width=65mm]{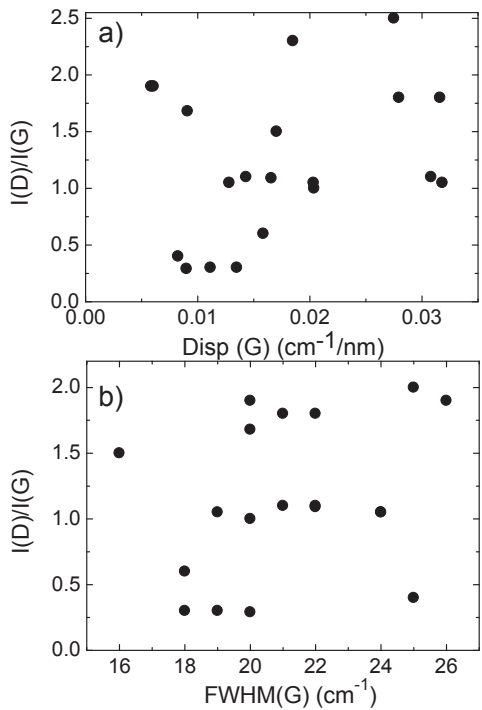}}
\caption{\label{raman2} a) I(D)/I(G) as function of Disp(G), b) I(D)/I(G) as function of FWHM(G) measured on flakes of our ink deposited on Si/SiO$_{2}$}
\end{figure}

We assign the D and D' peaks to the edges of the sub-micrometer flakes\cite{Casiraghi2009}, rather than to the presence of a large amount of disorder within the flakes. This is further supported by the plot of the G peak dispersion, Disp(G)(Fig.\ref{raman}b)). This is defined as Disp(G)\,=\,$\Delta Pos(G)/\Delta \lambda_{\rm L}$, where $\lambda_L$ is the laser excitation wavelength. Disp(G) is generated from the linear fit the plot of the G peak position, Pos(G), as a function of the laser excitation wavelength. In disordered carbons Pos(G) increases as the excitation wavelength decreases, from IR to UV\cite{Ferrari2000}, thus Disp(G) increases with disorder\cite{Ferrari2001,Ferrari2000}. The full width at half maximum of the G peak, FWHM(G), always increases with disorder\cite{Ferrari2003,Cancado2011}. Thus, combining the intensity ration of the D and G peaks, I(D)/I(G), with FWHM(G) and Disp(G) allows us to discriminate between disorder localized at the edges, and disorder in the bulk of the samples. In the latter case, to higher I(D)/I(G) would correspond higher FWHM(G) and Disp(G). Figs.\ref{raman2} a,b) show that Disp(G), I(D)/I(G) and FWHM(G) are not correlated, a clear indication that the major contribution to the D peak comes from the sample edges. Also, Disp(G) is nearly zero for all samples, compared to the values bigger than 0.1cm$^{-1}$/nm expected for disordered carbons\cite{Ferrari2001,Ferrari2004}, another indication of the lack of large structural disorder within our flakes. The distribution of 2D peak positions, Pos(2D), shown in \ref{raman}d, has two maxima$\sim$2692 and 2705cm$^{-1}$, similar to FWHM(2D) (\ref{raman}e). This is consistent with the samples being a distribution of SLG, BLG and FLGs, but with a significant fraction of SLGs. We note that for the flakes with the smallest Pos(2D) and FWHM(2D), the ration of the 2D and G integrated areas, A(2D)/A(G), is at most 3.5, implying a doping of at least 10$^{13}$cm$^{-2}$.\cite{Basko2009,Das2008,Pisana2007}

We now estimate $\eta$, $\rho$ and $\gamma$ for our ink, in order to check its viability for ink-jet printing. $\eta$ can be evaluated as $\eta=\eta_{0}$(1+2.5$\phi$)\cite{Kaye1997,Derby2003}, where $\eta_{0}$ is the viscosity of the pure solvent and $\phi$ the volume fraction of particles in the dispersion. We assume $\eta_{0}=\eta_{NMP}\sim$0.8mPas, the viscosity of pure NMP at $\sim$80$^\circ$C\cite{HandchemLide,Kauffman2001} (the temperature of the drops ejected from our printer, as specified in Ref. \onlinecite{Epson2011}). We take $\phi$=1-$\frac{Vol_{ink}}{Vol_{NMP}}$, where Vol$_{NMP}$ [$\sim$0.972 mm$^{3}$] is the volume of 1mg pure NMP and Vol$_{ink}$ [$\sim$0.94 mm$^{3}$] is the volume of 1mg of our ink, both measured by a micropipette ($\pm$2nL precision), at room temperature and pressure. We thus get $\phi\sim$0.03, and $\eta\sim$0.96mPas. From the same measurement we also obtain $\rho\sim$1.06gcm$^{-3}$ and derive $\gamma\sim$50mJ m$^{-2}$ from tensiometer measurements. Given these parameters, and our nozzle diameter$\sim$50$\mu$m, we get Z$\sim\frac{\sqrt{\gamma\rho a}}{\eta}\sim$1.7, which falls within the range suitable for printing\cite{Reis2000,Jang2009}, but close to the lower boundary of allowed Z,\cite{Fromm1984,Reis2000,Jang2009} thus implying a lower probability of secondary drops ejection\cite{Reis2000,Kaye1997,Derby2003}. However, high viscosity may generate nanoparticle re-aggregation\cite{Kaye1997}.
\subsection{Ink-jet printed features}
The final layout of printed nano-particle inks depends on substrate SE\cite{deGans2004,van Osch2008}, ink viscosity and surface tension\cite{deGans2004}.

To investigate the influence of surface treatments, we print our ink on pristine, HMDS coated and O$_{2}$ plasma treated Si/SiO$_{2}$. A modified Epson Stylus 1500 ink-jet printer equipped with an Epson S020049 cartridge is used to print the dispersions under a constant nitrogen flow, followed by annealing at 170$^\circ$C for 5 minutes to remove the NMP. The nozzle is placed$\sim$1mm above the substrate. HMDS is deposited by spin coating for 40s at 1000rpm, followed by annealing at 80$^\circ$C for 2 min. Alternatively the substrates are cleaned by a RF O$_{2}$ plasma at 200W and 4$\times$10$^{-1}$ Torr for 2 min.

\begin{figure}
\centerline{\includegraphics[width=85mm]{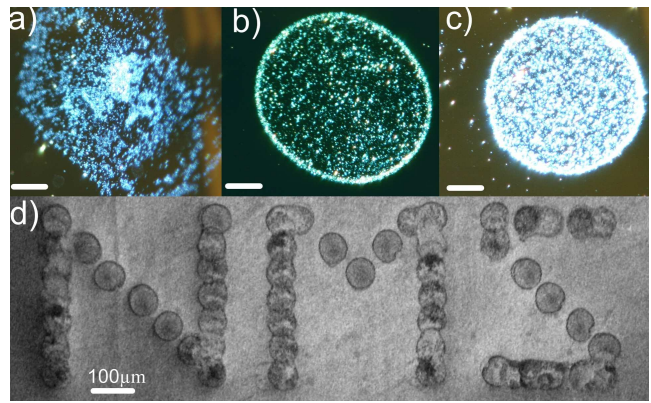}}
\caption{\label{drops} Dark field optical micrograph of inkjet printed drops on a) plasma cleaned, b) pristine and c) HMDS treated substrate. Scale: 20$\mu$m. d) SEM micrograph of printed pattern.}
\end{figure}

We use optical micrographs to visualize the ink-jet printed drops, Figs.\ref{drops}a,b,c. The bright green/blue color of the printed features is due to the use of dark field imaging. These reveal that HMDS constrains the drops to 90$\mu$m diameter (Fig.\ref{drops}c), smaller than on the other substrates ($\sim$100$\mu$m and $\sim$150$\mu$m for pristine, Fig.\ref{drops}b, and plasma treated SiO$_{2}$, Fig.\ref{drops}a). As discussed above, we use NMP as solvent to reduce the coffee ring effect compared to low boiling point solvents (e.g. water, chloroform)\cite{Deegan1997,Derby2003,Singh2010}. However, we still observe coffee-rings when printing on pristine SiO$_{2}$ (Fig.\ref{drops}b), while Fig.\ref{drops}c reveals a higher flake uniformity, and no coffee-rings on HMDS treated SiO$_{2}$. Fig.\ref{drops}d a representative printed pattern, showing the ability to fabricate complex layouts.
\begin{figure}
\centerline{\includegraphics[width=85mm]{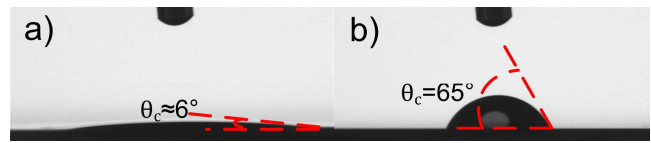}}
\caption{\label{contact_angle} Images of water drops dispensed on a) pristine and b) HMDS teated Si/SiO$_{2}$}
\end{figure}

Thus, HMDS appears to prevent coffee-rings. To understand this, we measure the substrates SE and investigate the printed stripes morphology, before and after surface treatment. We perform contact angle measurements with a A KSV CAM200 system. The contact angle is measured by dispensing 1$\mu$l DI water on the substrates. The surface tension is measured by the DuNouy-Padday technique\cite{Padday1971}. This consists in using a rod few millimeters diameter immersed in the dispersion, followed by pull out. The rod is attached to a scale or balance via a thin metal hook that measures the maximum pull force. This is recorded as the probe is first immersed 1mm into the solution and then slowly withdrawn from the interface. The contact angle, $\theta_{C}$, depends on the liquid surface tension\cite{Shafrin1960,deGennes1985,Israelachvili1991} and the substrate critical surface tension\cite{Shafrin1960,deGennes1985,Israelachvili1991}, according to the Young's relation\cite{deGennes1985,Israelachvili1991,Young1805}: $\gamma_{SV}$-$\gamma_{SL}$-$\gamma_{LV}$cos$\theta_{C}$=0, where $\gamma_{SV}$ [mJ m$^{-2}$] is the solid-vapor surface tension, $\gamma_{SL}$ is the solid-liquid surface tension and $\gamma_{LV}$ is the liquid-vapor surface tension.

Figs.\ref{contact_angle}a,b show ink drops printed onto pristine and HMDS treated Si/SiO$_{2}$, with $\theta_{C}\sim$6$^\circ$ and$\sim$65$^\circ$, indicating that the pristine substrate SE is modified following HMDS treatment. $\gamma_{LV}$ was measured$\sim$73mJ m$^{-2}$ in Ref.\onlinecite{Shafrin1967} for DI water, whereas $\gamma_{SV}\sim$116.5mJ m$^{-2}$ and$\sim$40mJ m$^{-2}$ were reported for pristine\cite{Thomas1996} and HMDS treated\cite{Glendinning1991} Si/SiO$_{2}$. Consequently, $\gamma_{SL}\sim$43.9mJ m$^{-2}$ and $\sim$9.1mJ m$^{-2}$ for pristine and HMDS treated Si/SiO$_{2}$, respectively. A higher $\gamma_{SL}$ implies a higher SE\cite{Ghatee2005}. Indeed, our $\gamma_{SL}$ correspond to SEs$\sim$73.9 and$\sim$39.1mJ m$^{-2}$ for pristine and HMDS treated Si/SiO$_{2}$. A small $\theta_{C}$ results in the drop rapid spreading on the substrate\cite{deGennes1985}, as seen in pristine SiO$_{2}$. On the other hand, HMDS provides higher $\theta_{C}$, since it lowers $\gamma_{SL}$ (thus the substrate SE), therefore reducing the wettability\cite{Marmur2003,Shafrin1960}.

When ink-jet printing stripes, the inter-drop (i.e. centre to centre) distance is an important parameter\cite{Duineveld2003}. For a large distance, individual drops are deposited on the substrate\cite{Duineveld2003,Reis2000,Derby2003}. As the inter-drop distance decreases, these merge into a line\cite{Duineveld2003}. Thus, in order to obtain a continuous line we need an inter-drop distance smaller than the drop diameter\cite{Duineveld2003}. On the other hand, Refs.\onlinecite{Kaye1997,Derby2003} reported that a very small inter-drop distance can result in particle aggregation on the substrate, thus a non-uniform stripe (i.e. irregular edges). We thus select an inter-drop distance suitable to have continuous lines, avoiding at the same time non-uniformities and irregular edges.

\begin{figure}
\centerline{\includegraphics[width=85mm]{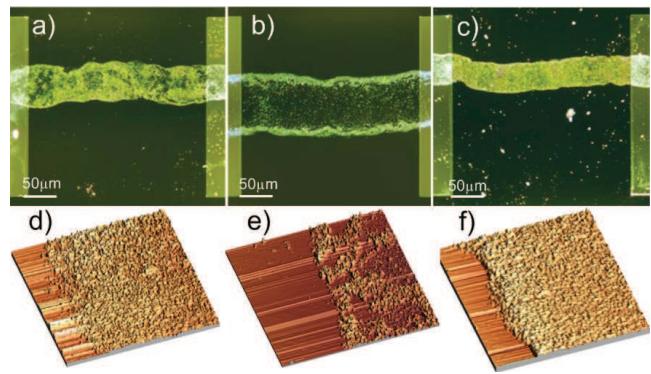}}
\caption{\label{stripes} Optical micrograph of ink-jet printed stripes on a) pristine, b) O$_{2}$ and c) HMDS treated substrates.d,e,f) AFM images of a,b,c}
\end{figure}

Figs.\ref{stripes}a,b,c are optical images of printed stripes on pristine, O$_{2}$ plasma treated and HMDS treated Si/SiO$_{2}$, whereas Figs.\ref{stripes}d,e,f plot the respective Atomic Force Microscope (AFM) topographies. The stripe in Fig.\ref{stripes}a is$\sim$100-110$\mu$m wide, has an average thickness$\sim$70nm and an irregular flake distribution, with aggregation of flakes. That in Fig.\ref{stripes}b is wider ($\sim$130-140$\mu$m), with aggregates at the edges, and an average thickness$\sim$55nm. The stripe in Fig.\ref{stripes}c has a more uniform and regular distribution of flakes,$\sim$85-90$\mu$m wide and$\sim$90nm thick. The width narrows going from the O$_{2}$ plasma treated to the HMDS treated Si/SiO$_{2}$, due to the SE decrease. Figs.\ref{stripes}d,e show stripes with voids and irregular flake distribution, with R$_{z}\sim$30-40nm. Fig.\ref{stripes}f presents a more homogeneous network with R$_{z}\sim$15nm. Thus, R$_{z}$ is lower when $\theta_{C}$ is higher, because the poor wettability of drops with higher $\theta_{C}$ reduces the stripe diameter (as shown in Figs.\ref{stripes}a,b,c), confining the flakes onto a smaller area. The uniformity of stripes printed on the HMDS treated substrate corroborates the above considerations on the SE changes. In fact, the presence of silane in HMDS\cite{Osthoff2007} promotes the adhesion of metallic particles to the substrate\cite{Osthoff2007,Gamerith2007}. Analogously, HMDS may promote the adhesion of graphene to the substrate, thus resulting in a uniform network.
\begin{figure}
\centerline{\includegraphics[width=90mm]{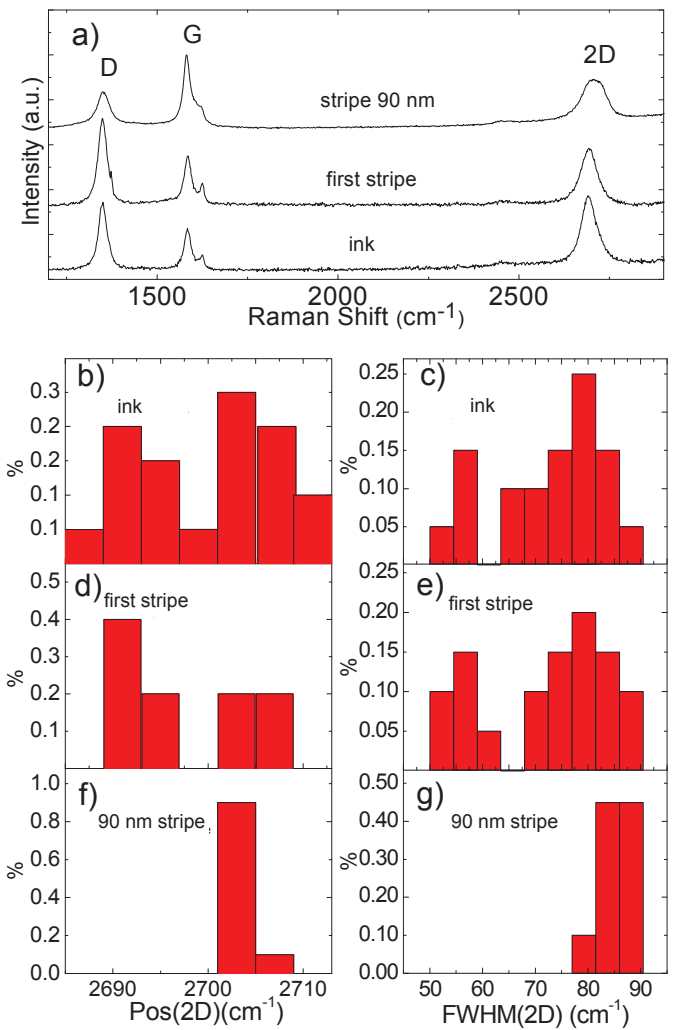}}
\caption{\label{Raman_stripes} a) Typical Raman spectrum of individual flakes in the ink, compared with spectra measured on the first stripe and on a stripe 90 nm thick. Pos(2D) and FWHM(2D) for b,c) ink; d,e) fist stripe; f,g) 90nm thick stripe}
\end{figure}
\begin{figure}
\centerline{\includegraphics[width=45mm]{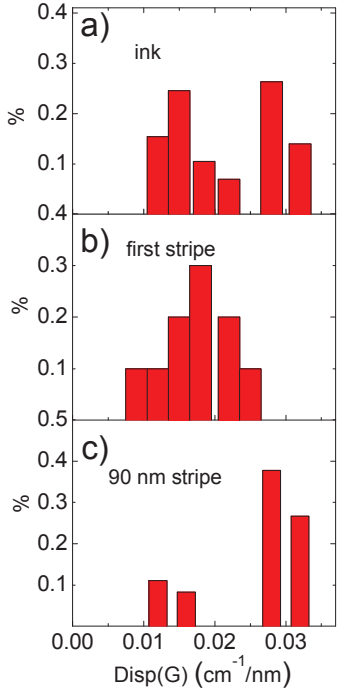}}
\caption{\label{Raman_stripes2} Distribution of Disp(G) for a) ink; b) fist stripe; c) 90nm thick stripe}
\end{figure}

Fig.\ref{Raman_stripes}a compares a typical Raman spectrum of a flake in the ink, with a measurement on the first stripe and on a stripe 90 nm thick, after 30 printing repetitions. Figs.\ref{Raman_stripes}b,c,d,e,f,g,\ref{Raman_stripes2} compare the Pos(2D), FWHM(2D) and Disp(G) distributions. The data show that the first stripe has very similar characteristics to the ink, as expected. However, the spectra after 90 repetitions show a Pos(2D) and FWHM(2D) distribution more typical of a multi-layer sample, having lost any direct signature of SLG. Note however that the 2D peak shape, even for the 90nm stripe, remains distinctly different from that of graphite. A similar aggregation of flakes was previously observed for thick films derived from graphene solutions\cite{Hernandez2008}. In all cases Disp(G) remains similar, and very low, again showing the lack of large amounts of defects within the flakes.
\subsection{Transparent and conductive patterns}
We now investigate the viability of our ink to print transparent and conductive patterns. We characterize the sheet resistance R$_{s}$ [$\Omega$/$\Box$] and Transmittance T [\%] of our stripes when placed on a transparent substrate. We thus use pristine, O$_{2}$ and HMDS treated borosilicate glass, with R$_{z}$$<$15nm similar to SiO$_{2}$ on Si, but with T$\sim$99\% (Pyrex 7740-Polished Prime Grade). T is measured on samples ink-jet printed on borosilicate glass (followed by annealing at 170$^\circ$C for 5 mins to remove the NMP) by scanning a 514.5nm laser beam with 100$\mu$m steps. The transmitted beam is measured with a photodiode. A microscope equipped with 100$\times$ long distance objective focuses the laser to$\sim$2$\mu$m. The incident power is kept at$\sim$8mW. The transmitted power is measured by a Ophir Nova II power meter with 0.1$\mu$W resolution.
\begin{figure*}
\centerline{\includegraphics[width=150mm]{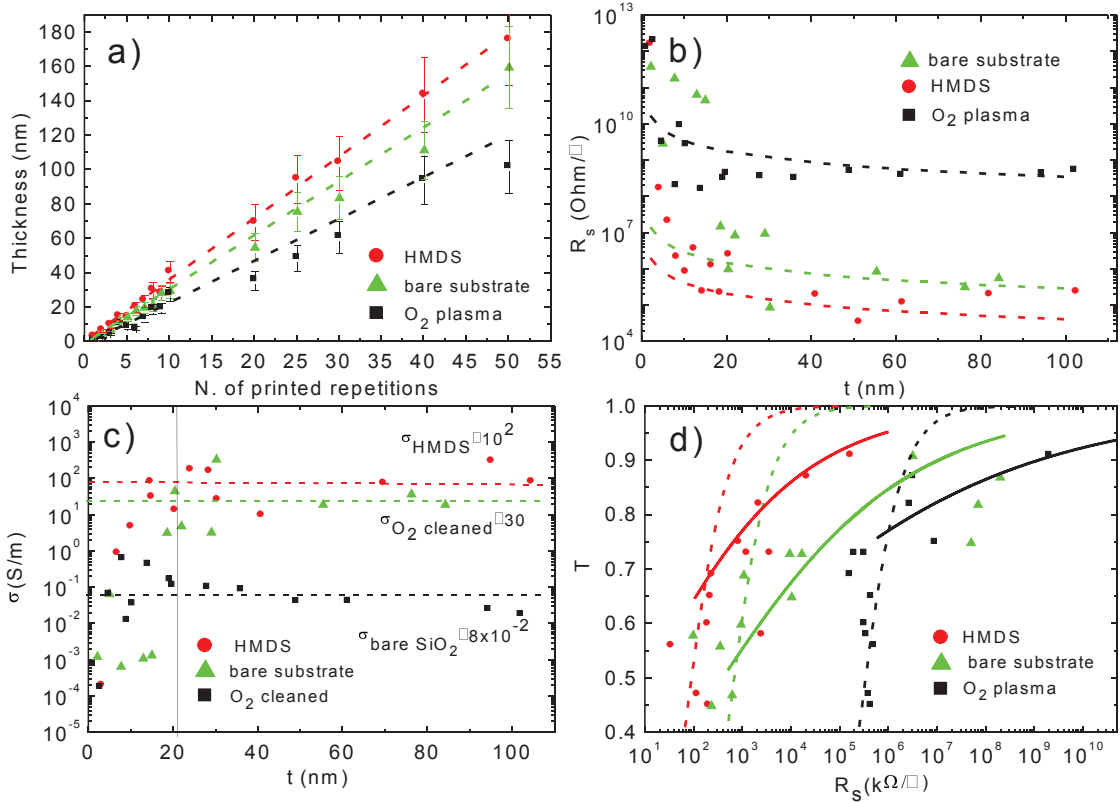}}
\caption{\label{sheet_resistance} a) Thickness as a function of printing repetitions. b, c) R$_{s}$ and $\sigma$ as a function of stripe thickness. d) T as a function of R$_{s}$ for HMDS coated (red dots), O$_{2}$ plasma treated (green triangles) and pristine (black squares) substrates}
\end{figure*}

Fig.\ref{sheet_resistance}a shows that for our stripes the experimentally measured thickness (t) increases linearly as a function of printing repetitions, with a slope defined by the surface treatment. Fig.\ref{sheet_resistance}b plots the four-probe measured R$_{s}$ as a function of t. For large t, R$_{s}$ settles to$\sim$34,$\sim$500,$\sim$10$^{5}$k$\Omega$/$\Box$ for HMDS treated, pristine and O$_{2}$ treated glass, respectively. For t$<$20nm, R$_{s}$ increases for all substrates. For a thin film, R$_{s}=(\sigma~t)^{-1}$, where $\sigma$ [S/m] its conductivity\cite{Smits1958}. Thus, from Fig.\ref{sheet_resistance}b and $\sigma$=(R$_{s}~t)^{-1}$, we get the data in Fig.\ref{sheet_resistance}c. $\sigma$ is constant for t$>$20nm, in the case of HMDS treated, pristine and plasma treated glass, with an average$\sim$10$^{2}$,$\sim$30,$\sim$10$^{-1}$S/m, respectively. Thus, stripes on HMDS treated glass have an higher $\sigma$ combined with a more regular network of flakes, compared to the other two substrates. When t$<$20nm, $\sigma$ decreases for all substrates. A similar trend was reported for CNT films on SiO$_{2}$ (produced by vacuum filtration)\cite{Hu2004,Geng2007}, ink-jet printed CNT patterns on SiO$_{2}$\cite{Takenobu2009,Okimoto2010}, graphene films on SiO$_{2}$,\cite{DeSmall2009,DeACS2010} and Polyethylene-terephthalate(PET),\cite{DeSmall2009,DeACS2010} as well as Ag nanowire films, produced by vacuum filtration on SiO$_{2}$\cite{DeACS2010}. Refs. \onlinecite{Hu2004,Geng2007,DeSmall2009,DeACS2010} explained this decrease of $\sigma$ for small t, due to percolation.

The percolation theory\cite{Kirkpatrick1973} predicts $\sigma$, for a network of conductive particles, to scale as\cite{Kirkpatrick1973}:
\begin{equation}
\label{percolation1}
\sigma\propto (X-X_{c})^\beta
\end{equation}
where $X$ [$\mu$g/mm$^{2}$] is the concentration of conductive particles per unit area, $X_{c}$ [$\mu$g/mm$^{2}$] is the critical concentration of flakes corresponding to the percolation threshold and $\beta$ is the percolation exponent. Eq.\ref{percolation1} can be rewritten in terms of t, rather than $X$\cite{Hu2004} as:
\begin{equation}
\label{percolation2}
\sigma\propto (t-t_{c})^\epsilon
\end{equation}
where $t_{c}$ is the critical thickness and $\epsilon$ is the percolation exponent. Fig.\ref{sheet_resistance}c shows two regimes for $\sigma$ as a function of t: a percolative linear behavior for t$<$20nm and a constant $\sigma_{bulk}$ for t$>$20nm. This can be explained considering that our films stop behaving like bulk materials below a critical thickness ($t_{min}$).

\begin{figure}
\centerline{\includegraphics[width=85mm]{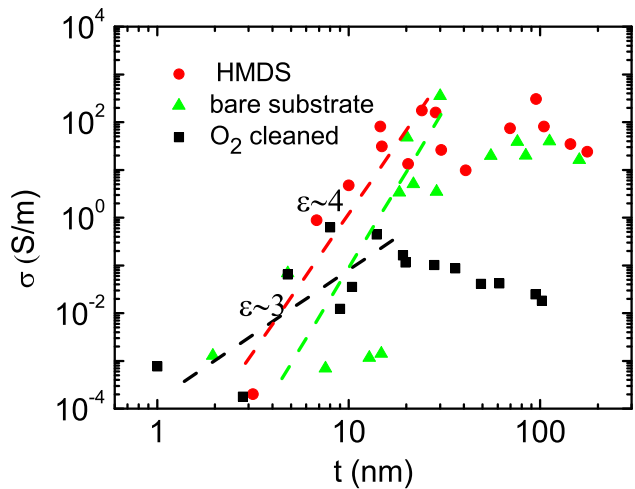}}
\caption{\label{percolation_coeff} Conductivity as a function of thickness, in logarithmic scale, for stripes printed on HMDS treated (red dots), O$_{2}$ treated (green triangles) and pristine (black squares) substrates. Lines are fits in the percolation regime}
\end{figure}

The exponent $\epsilon$ can be estimated by a linear fit of the log$_{10}$ plot of $\sigma$ vs t, in the percolation region (t$<$20nm), Fig.\ref{percolation_coeff}. We get $\epsilon$$\sim$4 for stripes on HMDS treated and pristine glass, while $\epsilon\sim$3 for O$_{2}$ treated glass. These values indicate percolation, as reported by Refs.\onlinecite{Kogut1979,Stauffer1985,Johner2008,DeACS2010} for networks with various geometries. $\epsilon$ is expected to increase with particle size\cite{Kogut1979,Johner2008} and decrease with $X_{c}$\cite{Kogut1979,Johner2008}. Assuming a similar particle size, since the same ink is used for all cases, we deduce that $\epsilon$$\sim$4 points to a bigger $X_{c}$ than $\epsilon\sim$3. This indicates formation of a more uniform network on HMDS treated and pristine glass compared to O$_{2}$ treated glass.

We also determine the minimum concentration necessary to achieve the bulk conductivity regime. To do so, we assume X$\gg$X$_{C}$, because the bulk regime needs a tight network of interconnected flakes\cite{Stauffer1985,Doherty2009,DeACS2010}. Given our $c\sim$0.11g/L, volume per printed drop$\sim$10nL\cite{Epson2011}, and a dried drop size on the three substrates of$\sim$90,100,130$\mu$m, we estimate $X\sim$4$\times$10$^{-2}$,$\sim$10$^{-2}$ and$\sim$0.7$\times$10$^{-2}\mu$g/mm$^{2}$ for stripes printed on HMDS, pristine and plasma treated glass, respectively. Consequently, from Eq.\ref{percolation1}, $\sigma$ for stripes on HMDS treated glass ($\sim$10$^{2}$S/m) is higher than on pristine ($\sim$40S/m) and plasma treated glass($\sim$0.1S/m).

Fig.\ref{sheet_resistance}d shows T as a function of R$_{s}$. The dashed lines are a plot of the relation T=$\left(1+\frac{Z_{0}~G_{0}}{2R_{s}\sigma_{bulk}}\right)^{-2}$ expected for stripes with $\sigma_{bulk}$ conductivity, where Z$_{0}$=377$\Omega$ is the free-space impedance, G$_{0}\sim$6$\times$10$^{-5}\Omega^{-1}$ is the universal optical conductance of graphene\cite{Kuzmenko2008}. The solid lines are a plot of T=$\left[1+\frac{1}{\Pi}~\left(\frac{Z_{0}}{R_{s}}\right)^{1/(\epsilon+1)}\right]^{-2}$ expected in the percolative regime\cite{DeACS2010}, where $\Pi$ is the percolative Figure of Merit $\Pi=2\left[\frac{\sigma_{bulk}/G_{0}}{(Z_{0}~t_{min}~G_{0})^{\epsilon}}\right]^{1/(\epsilon+1)}$. Our experimental T deviates from the dashed lines for T$>$75\%. We assign this to the percolative regime where $\sigma_{DC}$ deviates from a bulk-like behavior. Also in this case, printing on HMDS treated glass gives the highest T for a given R$_{s}$.
\subsection{Ink jet printed devices}
Ink-jet printed TFTs based on organic semiconducting polymers have been widely investigated\cite{Sirringhaus2000,Ong2004,Arias2004}. The current state of the art devices have $\mu$ ranging from 0.01 to$\sim$0.5cm$^{2}$V$^{-1}$s$^{-1}$, with ON/OFF ratios up to 10$^{5}$.\cite{Ong2004,Arias2004,Wu2005}
Several Inkjet printed TFTs using various carbon nanomaterials have been reported. For example, fullerene-based TFTs were discussed in Refs. \onlinecite{Kaneto2007,Morita2010}, with $\mu$ up to 0.01cm$^{2}$V$^{-1}$s$^{-1}$ and an ON/OFF ratio$<$10. TFTs printed from CNT-based inks have been presented by several groups\cite{Beecher2007,Hsieh2009,Takenobu2009,Okimoto2009,Ha2010}. The highest $\mu$ thus far is$\sim$50cm$^{2}$V$^{-1}$s$^{-1}$ combined with an ON/OFF ratio 10$^{3}$, but measured at 10$^{-6}$ Torr and 100K\cite{Ha2010}. Ink-jet printed TFTs from GO-based inks were discussed in Refs. \onlinecite{Dua2010,Wang2009}, with $\mu$ up to$\sim$90cm$^{2}$V$^{-1}$s$^{-1}$ for an ON/OFF ratio of 10 (measured at room conditions), after GO reduction.
\begin{figure}
\centerline{\includegraphics[width=90mm]{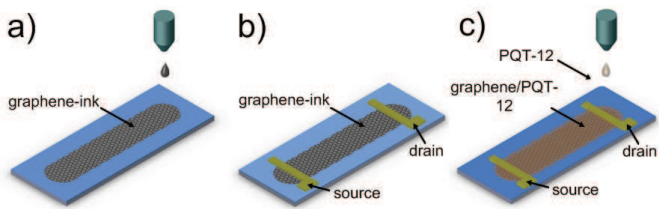}}
\caption{\label{TFT_schem} a) Ink on Si/SiO$_{2}$.b) Cr-Au pads define the source and drain contacts. c) A layer of Poly[5,5'-bis(3-dodecyl-2-thienyl)-2,2'-bithiophene] (PQT-12) is printed on top}
\end{figure}

We print our TFTs as for Fig.\ref{TFT_schem}a, and contact them with chromium-gold source and drain pads (Fig.\ref{TFT_schem}b). The transfer characteristics are measured (at room conditions) at different drain voltages (V$_{d}$=-2,-4,-8V). $\mu$ is derived from $\mu$=$\frac{L}{W\;C_{i}\;V_{d}}\;\frac{dI_{d}}{dV_{g}}$, where L [$\mu$m] and W [$\mu$m] are the channel length and width respectively, C$_{i}$ is the gate dielectric capacitance ($\sim$10nF/cm$^{2}$)\cite{Oh2009}. We get $\mu\sim$95cm$^{2}$V$^{-1}$s$^{-1}$ for an ON/OFF ratio$\sim$10 at V$_{d}$=-2V, comparable to that reported in Ref. \onlinecite{Wang2009} for ink-jet printed RGO TFTs. $\mu$ in our devices is almost four orders of magnitude higher than printed fullerene-based TFTs\cite{Kaneto2007,Morita2010} (for the same ON/OFF ratio) and more than two orders higher than ink-jet printed CNTs\cite{Beecher2007,Takenobu2009} (for a ON/OFF ratio of 10). However, the ON/OFF ratio in our TFTs is lower than the state of the art for CNTs (but measured at 10$^{-6}$ Torr and 100K) at similar $\mu$\cite{Ha2010}. We note that ink-jet printed electronics requires high $\mu$ at room conditions\cite{Singh2010,Singh2006}. So far CNT ink-jet printed devices measured at room conditions have $\mu$ no larger than $\sim$1cm$^{2}$V$^{-1}$s$^{-1}$ (at ON/OFF$\sim$10)\cite{Takenobu2009}, which is two orders of magnitude smaller than our jet printed TFTs.
\begin{figure}
 \centerline{\includegraphics[width=60mm]{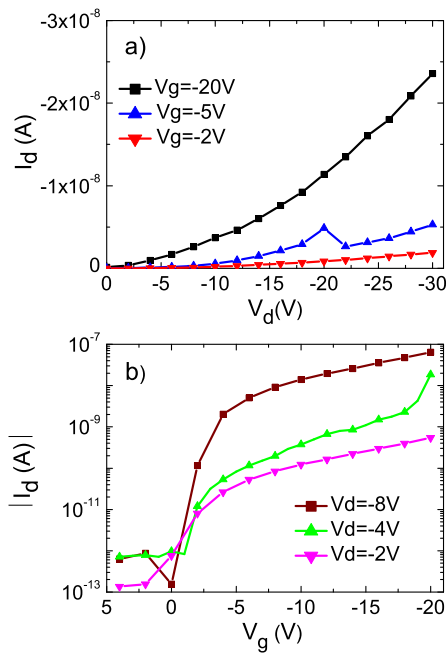}}
 \caption{\label{TFT_charact} a) Output and b) transfer characteristics of an ink-jet printed graphene/PQT TFT.}
\end{figure}

Organic semiconducting inks\cite{Ong2004,Arias2004,Wu2005} suffer from low $\mu$, limited by variable range hopping between the isolated polymer chains\cite{Sirringhaus1998}. The overall charge conduction in crystalline organic semiconducting thin films is determined by both intra-chain and inter-chain charge transport\cite{Song2010}. The former is much faster than inter-chain hopping\cite{Song2010, Sirringhaus1998}. Many groups have tried to improve interchain hopping\cite{Whiting2009,Klauk_book2006,Beecher2007,Hsieh2009}. Ref. \onlinecite{Whiting2009} proposed a chemical modification of the semiconducting organic ink by electron acceptors, while embedding Au nano-particles in the semiconducting organic ink was proposed by Ref. \onlinecite{Klauk_book2006}. Embedding CNTs in the semiconducting ink\cite{Beecher2007,Hsieh2009} allowed us to get $\mu \sim0.07cm^{2}$V$^{-1}$s$^{-1}$ at room conditions.

We combine our graphene-ink with one of the most commonly used organic polymers for ink-jet printing, Poly[5,5'-bis(3-dodecyl-2-thienyl)-2,2'-bithiophene] (PQT-12)\cite{Ong2004,Arias2004,Wu2005} in order to investigate its viability as interchain hopping enhancer (similarly to Au nanoparticles and CNTs). PQT-12 is widely used due to the higher environmental stability (up to 300 days at room conditions\cite{Chason2005}), with respect to other organic semiconducting inks\cite{Klauk_book2006,Chason2005}. Graphene can bridge the polymer chains, allowing a more efficient charge transport.

We fabricate a graphene/PQT-12 TFT following the steps shown in Figs.\ref{TFT_schem}a,b,c. Fig.\ref{TFT_charact}a plots its output characteristics at V$_{g}$=-2,-5,-20 V. For each V$_{g}$, V$_{d}$ is swept from 0 to -30 V in steps of 2V. At V$_{d}$=-2V, we get $\mu\sim$0.17cm$^{2}$V$^{-1}$s$^{-1}$ and an ON/OFF ratio$\sim$4$\times$10$^{5}$. This $\mu$ is about ten times that of ink-jet printed CNTs/PQT-12 TFTs\cite{Beecher2007,Hsieh2009} at ON/OFF$\sim$10$^{5}$. When compared to pure organic semiconducting polymers, our $\mu$ is $\sim$20 times higher than ink-jet printed PQT-12\cite{Arias2004,Wu2005}, and twice the highest reported $\mu$ for ink-jet printed TFTs made of pure (Poly(2,5-bis(3-tetradecyllthiophen-2-yl)thieno[3,2-b]thiophene)\cite{Kawase2005,Klauk_book2006,Parmer2008,Singh2010}. Thus, the combination of our graphene-ink with organic semiconducting inks is promising for high performance printed electronics.
\section{Conclusions}
We demonstrated ink-jet printing of graphene. Liquid phase exfoliated graphene is an ideal and low cost material for the fabrication of transparent conductive inks. Our graphene-ink was used to print TFTs with $\mu$ up to$\sim$95cm$^{2}$V$^{-1}$s$^{-1}$. It was also combined with PQT-12 to fabricate devices with $\mu \sim$0.2cm$^{2}$V$^{-1}$s$^{-1}$ and ON/OFF ratios$\sim$4$\times$10$^{5}$. This demonstrates the viability of graphene-inks for flexible and transparent electronics.
\section{acknowledgements}
We acknowledge funding from the Royal Society Brian Mercer Award for Innovation, the ERC grant NANOPOTS, EPSRC grants EP/GO30480/1 and EP/F00897X/1, EU Grants RODIN and GENIUS, King's college, Cambridge. ACF is a Royal Society Wolfson Research Merit Award holder.


\begin{thebibliography}{150}

\bibitem{Cao2008}
Q. Cao, H. S. Kim, N. Pimparkar, J. P. Kulkarni, C. J. Wang, M. Shim, K. Roy, M. A. Alam, J. A. Rogers, Nature \textbf{454}, 495 (2008).

\bibitem{Zhou2006}
L. Zhou, A. Wanga, S.C. Wu, J. Sun, S. Park, T. N. Jackson, Appl. Phys. Lett. \textbf{88},  083502 (2006).

\bibitem{Ota1973}
I. Ota, J. Ohnishi and M. Yoshiyama, M. Proc. IEEE \textbf{61}, 832 (1973).

\bibitem{Gelinck2004}
G. H. Gelinck, H. E. A. Huitema, E. van Veenendaal, E. Cantatore, L. Schrijnemakers, J. B. P. H. van der Putten, T. C. T. Geuns, M. Beenhakkers, J. B. Giesbers, B.-H. Huisman, E. J. Meijer, E. M. Benito, F. J. Touwslager, A. W. Marsman, B. J. E. van Rens, D. M. de Leeuw, Nat. Mater. \textbf{3}, 106 (2004).

\bibitem{Sekitani2009}
T. Sekitani, T. Yokota, U. Zschieschang, H. Klauk, S. Bauer, K. Takeuchi, M. Takamiya, T. Sakurai, T. Someya, Science \textbf{326}, 1516 (2009).

\bibitem{Myny2009}
K. Myny, S. Steudel, P. Vicca, M. J. Beenhakkers, N. A. J. M. van Aerle, G. H. Gelinck, J. Genoe, W. Dehaene, P. Heremans, Solid State Electron. \textbf{53}, 1220 (2009).

\bibitem{Granqvist2007}
C. G., Granqvist, Sol. Energ. Mat. Sol. C. \textbf{91}, 1529 (2007).

\bibitem{Yoon2008}
J. Yoon, A. J. Baca, S.-I. Park, P. Elvikis, J. B. Geddes, L. Li, R. H. Kim, J. Xiao, S. Wang, T.-H. Kim, M. J. Motala, B. Y. Ahn, E. B. Duoss, J. A. Lewis, R. G. Nuzzo, P. M. Ferreira, Y. Huang, A. Rockett, J. A. Rogers, Nat. Mater. \textbf{7},  907 (2008).

\bibitem{Schmied2009}
B. Schmied, J. Gunther, C. Klatt, H. Kober, E. Raemaekers, Smart Textiles \textbf{60}, 67 (2009).

\bibitem{kim2008}
D. Kim, A. Jong-Hyun, K. Hoon-Sik, L. Keon Jae, K. Tae-Ho, Y. Chang-Jae, R. G. Nuzzo, J. A.  Rogers, IEEE Electr. Device Lett. \textbf{29}, 73 (2008).

\bibitem{Singh2006}
T. B. Singh, N. S. Sariciftci, Annu. Rev. Mater. Res. \textbf{36}, 199 (2006).

\bibitem{Rogers2001}
J. A. Rogers, Z. Bao, K. Baldwin, A. Dodabalapur, B. Crone, V. R. Raju, V. Kuck, H. Katz, K. Amundson, J. Ewing, P. Drzaic, P. Natl. Acad. Sci. U.S.A. \textbf{98}, 4835 (2001).

\bibitem{Forrest2004}
S. R. Forrest, Nature \textbf{428}, 911 (2004).

\bibitem{Bao1999}
Z. Bao, J. A. Rogers, H. E. Katz, J. Mater. Chem. \textbf{9}, 1895(1999).

\bibitem{Sirringhaus2000}
H. Sirringhaus, T. Kawase, R. H. Friend, T. Shimoda, M. Inbasekaran, W. Wu, E. P. Woo, Science \textbf{290}, 2123 (2000).

\bibitem{Sun2006}
Y. G. Sun, E. Menard, J. A. Rogers, H. S. Kim, S. Kim, G. Chen, I. Adesida, R. Dettmer, R. Cortez, A. Tewksbury, Appl. Phys. Lett. \textbf{88}, 3 (2006).

\bibitem{McAlpine2005}
M. C. McAlpine, R. S. Friedman, C. M. Lieber, Proc. IEEE \textbf{93}, 1357 (2005).

\bibitem{Singh2010}
M. Singh, H. M. Haverinen, P. Dhagat, G. E. Jabbour, Adv. Mater. \textbf{22}, 673 (2010).

\bibitem{Peumans2003}
P. Peumans, S. Uchida, S. R. Forrest, Nature \textbf{425}, 158 (2003).

\bibitem{Servati2005}
P. Servati, A. Nathan, Proc. IEEE \textbf{93}, 1257(2005).

\bibitem{deGans2004}
B. J., DeGans,  P. Duineveld, U. Schubert, Adv. Mater. \textbf{16}, 203 (2004).

\bibitem{Dong2006}
H. M. Dong, W. W. Carr, J. F. Morris, Phys. Fluids \textbf{18}, 16 (2006).

\bibitem{van Osch2008}
T. H. J. van Osch, J. Perelaer, A. W. M. de Laat, U. S. Schubert,  Adv. Mater. \textbf{20}, 343 (2008).

\bibitem{Yoo2010}
J. E. Yoo, K. S. Lee, A. Garcia, J. Tarver, E. D. Gomez, K. Baldwin, Y. Sun, H. Meng, T.Q. Nguyen, Y.L. Loo, Proc. Natl. Acad. Sci. U.S.A. \textbf{107}, 5712 (2010).

\bibitem{Shimoda2006}
T. Shimoda, Y. Matsuki, M. Furusawa, T. Aoki, I. Yudasaka, H. Tanaka, H. Iwasawa, D. Wang, M. Miyasaka, Y. Takeuchi, Nature \textbf{440}, 783 (2006).

\bibitem{Noh2007}
Y. Y. Noh, X. Cheng, H. Sirringhaus, J. I. Sohn, M. E. Welland, D. J. Kang, Appl. Phys. Lett. \textbf{91}, 043109 (2007).

\bibitem{Beecher2007}
P. Beecher, P. Servati, A. Rozhin, A. Colli, V. Scardaci, S. Pisana, T. Hasan, A. J. Flewitt, J. Robertson, G. W. Hsieh, F. M. Li, A. Nathan, A. C. Ferrari, W. I. Milne, J. Appl. Phys. \textbf{102}, 043710 (2007).

\bibitem{Hsieh2009}
G. W. Hsieh, F. M. Li, P. Beecher, A. Nathan, Y. L. Wu, B. S. Ong, W. I. Milne, J. Appl. Phys. \textbf{106}, 7 (2009).

\bibitem{Takenobu2009}
T., Takenobu, N. Miura, S.Y. Lu, H. Okimoto, T. Asano, M. Shiraishi, Y. A. Iwasa, App. Phys. Expr. \textbf{2}, 025005 (2009).

\bibitem{Okimoto2010}
H. Okimoto, T. Takenobu, K. Yanagi, Y. Miyata, H. Shimotani, H. Kataura, Y. Iwasa, Adv. Mater. \textbf{22}, 3981(2010).

\bibitem{Okimoto2009}
H. Okimoto, T. Takenobu, K. Yanagi, Y. Miyata, H. Kataura, T. Asano, Y. Iwasa, J. J. App. Phys. \textbf{48}, 4 (2009).

\bibitem{Ha2010}
M. Ha, Y. Xia, A. A. Green, W. Zhang, M. J. Renn, C. H. Kim, M. C. Hersam, C. D. Frisbie, ACS Nano \textbf{4}, 4388 (2010).

\bibitem{Luechinger2008}
N. Luechinger, A., E. K. Athanassiou, W. J. Stark, Nanotechnol. \textbf{19}, 445201 (2008).

\bibitem{Geim_rise2007}
A. K. Geim, K. S. Novoselov, Nat. Mater. \textbf{6}, 183 (2007).

\bibitem{Novoselov2004}
K. S. Novoselov, A. K. Geim, S. V. Morozov, D. Jiang, Y. Zhang, S. V. Dubonos, I. V. Grigorieva, A. A. Firsov, Science \textbf{306}, 666 (2004).

\bibitem{Charlier2008}
J. C. Charlier, P. C. Eklund, J. Zhu, A. C. Ferrari, Topics Appl. Phys. \textbf{111}, 673 (2008).

\bibitem{Bonaccorso2010}
F. Bonaccorso, Z. Sun, T. Hasan, A. C. Ferrari, Nat. Photon. \textbf{4}, 611 (2010).

\bibitem{Lin2010}
Y. M. Lin, C. Dimitrakopoulos, K. A. Jenkins, D. B. Farmer, H. Y. Chiu, A. Grill, P. Avouris, Science \textbf{327}, 662 (2010).

\bibitem{sunacsnano10}
Z. Sun, T. Hasan, F. Torrisi, D. Popa, G. Privitera, F. Wang, F. Bonaccorso, D. M. Basko, A.C. Ferrari, ACS Nano \textbf{4}, 803 (2010).

\bibitem{NovoselovPNAS2005}
K. S. Novoselov, D. Jiang, F. Schedin, T. J. Booth, V. V. Khotkevich, S. V. Morozov, A. K. Geim, PNAS \textbf{102}, 10451 (2005).

\bibitem{Karu1966}
A. E. Karu,M. Beer, J. Appl. Phys. \textbf{37}, 2179 (1966).

\bibitem{Obraztsov2007}
A. N. Obraztsov, E. A. Obraztsova, A. V. Tyurnina, A. A. Zolotukhin, Carbon \textbf{45}, 2017 (2007).

\bibitem{Kim2009}
K. S. Kim, Y. Zhao, H. Jang, S. Y. Lee, J. M. Kim, K. S. Kim, J.-H. Ahn, P. Kim, J.Y. Choi, B. H. Hong, Nature \textbf{457}, 706 (2009).

\bibitem{Reina2009}
A. Reina, X. Jia, J. Ho, D. Nezich, H. Son, V. Bulovic, M. S. Dresselhaus, J. Kong, Nano Lett. \textbf{9}, 30 (2009).

\bibitem{Li2009}
X. S. Li, W. W. Cai, J. H. An, S. Kim, J. Nah, D. X. Yang, R. Piner, A. Velamakanni, I. Jung, E. Tutuc, S. K. Banerjee, L. Colombo, R.S. Ruoff, Science \textbf{324}, 1312 (2009).

\bibitem{Bae2010}
S. Bae, H. Kim, Y. Lee, X. Xu, J.-S. Park, Y. Zheng, J. Balakrishnan, T. Lei, H. Ri Kim, Y. I. Song, Y.J. Kim, K. S. Kim, B. Ozyilmaz, J.H. Ahn, B. H. Hong,S. Iijima, Nat. Nano. \textbf{5}, 574 (2010).

\bibitem{Berger2006}
C. Berger, Z. M. Song, X. B. Li, X. S. Wu, N. Brown, C. Naud, D. Mayou, T. B. Li, J. Hass, A. N. Marchenkov, E. H. Conrad, P. N. First,W. A. de Heer, J. Phys. Chem. B \textbf{108}, 19912 (2006).

\bibitem{Acheson1896}
E. G. Acheson, US patent \ 615 (1896).

\bibitem{Badami1962}
D. V. Badami, Nature \textbf{193}, 569 (1962).

\bibitem{Emtsev2009}
K. V. Emtsev, A. Bostwick, K. Horn, J. Jobst, G. L. Kellogg, L. Ley, J. L. McChesney, T. Ohta, S. A. Reshanov, J. Rohrl, E. Rotenberg, A. K. Schmid, D. Waldmann, H. B. Weber,T. Seyller, Nat. Mater. \textbf{8}, 203 (2009).

\bibitem{Oshima1997}
C. Oshima, A. Nagashima, J. Phys.: Condens. Matter \textbf{9}, 1 (1997).

\bibitem{Gamo1997}
Y. Gamo, A. Nagashima, M. Wakabayashi, M. Terai, C. Oshima, Surf. Sci. \textbf{374}, 61 (1997).

\bibitem{Rosei1983}
R. Rosei, M. De Crescenzi, F. Sette, C. Quaresima, A. Savoia, P. Perfetti, Phys. Rev. B \textbf{28}, 1161 (1983).

\bibitem{Sutter2008}
P. W. Sutter, J. I. Flege, E. A. Sutter, Nat. Mater. \textbf{7}, 406 (2008).

\bibitem{Hernandez2008}
Y. Hernandez, V. Nicolosi, M. Lotya, F. M. Blighe, Z. Sun, S. De, I. T. McGovern, B. Holland, M. Byrne, Y. K. Gun'Ko, J. J. Boland, P. Niraj, G. Duesberg, S. Krishnamurthy, R. Goodhue, J. Hutchison, V. Scardaci, A. C. Ferrari, J. N. Coleman, Nat. Nanotech. \textbf{3}, 563 (2008).

\bibitem{Lotya2009}
M. Lotya, Y. Hernandez, P. J. King, R. J. Smith, V. Nicolosi, L. S. Karlsson, F. M. Blighe, S. De, Z. Wang, I. T. McGovern, G. S. Duesberg, J. N. Coleman, J. Am. Chem. Soc. \textbf{131}, 3611 (2009).

\bibitem{Valles2008}
C. Valles, C. Drummond, H. Saadaoui, C. A. Furtado, M. He, O. Roubeau, L. Ortolani, M. Monthioux and A. Penicaud, J. Am. Chem. Soc \textbf{130}, 15802 (2008).

\bibitem{Hasan2010}
T. Hasan, F. Torrisi, Z. Sun, D. Popa, V. Nicolosi, G. Privitera, F. Bonaccorso, A. C. Ferrari, Phys. Stat. Sol. B \textbf{247}, 2953 (2010).

\bibitem{Marago2010}
O. M. Marago, P. H. Jones, F. Bonaccorso, V. Scardaci, P. G. Gucciardi, A. G. Rozhin, A. C. Ferrari, ACS Nano \textbf{4}, 7515 (2010).

\bibitem{Greennlett2009}
A. A. Green, M. C. Hersam, Nano Lett. \textbf{9}, 4031 (2009).

\bibitem{LiScience2008}
X. L. Li, X. R. Wang, L. Zhang, S. W. Lee, H. J. Dai, Science \textbf{319}, 1229 (2008).

\bibitem{Stankovich2006}
S. Stankovich, R. D. Piner, S. T. Nguyen, R. S. Ruoff, Carbon \textbf{44}, 3342 (2006).

\bibitem{Hummers1958}
W. S. Hummers, R. E. Offeman, J. Am. Chem. Soc. \textbf{80}, 1339 (1958).

\bibitem{Brodie1860}
B. C. Brodie, Ann. Chim. Phys. \textbf{59}, 466 (1860).

\bibitem{Staudenmaier1898}
L. Staudenmaier,  Ber. Deut. chem. Ges. \textbf{31}, 1481 (1898).

\bibitem{Mattevi2009}
C. Mattevi, G. Eda, S. Agnoli, S. Miller, K. A. Mkhoyan, O. Celik, D. Mastrogiovanni, G. Granozzi, E. Garfunkel, M. Chhowalla, Adv. Funct. Mater. \textbf{29}, 2577 (2009).

\bibitem{Cai2008}
W. W. Cai, R. D. Piner, F. J. Stadermann, S. Park, M. A. Shaibat, Y. Ishii, D. X. Yang, A. Velamakanni, S. J. An, M. Stoller, J. H. An, D. M. Chen, R. S. Ruoff,  Science \textbf{321}, 1815 (2008).

\bibitem{Eda2010}
G. Eda, M. Chhowalla, Adv. Mater. \textbf{22}, 2392 (2010).

\bibitem{GO_organic_08}%
J. I., Paredes, S. Villar-Rodil, A. Martinez-Alonso, J. M. D. Tascon, Langmuir, \textbf{24}, 10560 (2008).

\bibitem{he_GO_model_98}%
H., He, J. Klinowski, M. Forster, A. Lerf, Chem. Phys. Lett., \textbf{287}, 53 (1998).

\bibitem{Eda2008}
G. Eda, G. Fanchini, M. Chhowalla, Nat. Nanotech. \textbf{3}, 270 (2008).

\bibitem{Dua2010}
V. Dua, S. Surwade, S. Ammu, S. Agnihotra, S. Jain, K. Roberts, S. Park, R. Ruoff, S. Manohar, Angew. Chem. Int. Ed. \textbf{49}, 2154 (2010).

\bibitem{Wang2009}
S. Wang, P. K. Ang, Z. Wang, A. L. L. Tang, J. T. L. Thong, K. P. Loh, Nano Lett. \textbf{10}, 92 (2009).

\bibitem{Park2007}
B. K. Park, D. Kim, S. Jeong, J. Moon, J. S. Kim, Thin Solid Films \textbf{515}, 7706 (2007).

\bibitem{Reis2000}
N. Reis, B. Derby, MRS. Symp. Proc. \textbf{624}, 65 (2000).

\bibitem{Jang2009}
D. Jang, D. Kim and J. Moon, Langmuir \textbf{25}, 2629 (2009).

\bibitem{Fromm1984}
J. E. Fromm, IBM J. Res. Dev., \textbf{28}, 322 (1984).

\bibitem{Microfab1999}
http://www.microfab.com/equipment/technotes/technote99-02.pdf.

\bibitem{deGennes1985}
P. G. De Gennes, Rev. Mod. Phys. \textbf{57}, 827 (1985).

\bibitem{Shafrin1960}
E. G. Shafrin, W. A. Zisman, J. Phys. Chem. \textbf{64}, 519 (1960).

\bibitem{Israelachvili1991}
J. Israelachvili, \textit{Intermolecular and Surface Forces}; Academic press, New York, (1991).

\bibitem{Derby2003}
B. Derby, N. Reis, MRS. Bull. \textbf{28}, 815 (2003).

\bibitem{Park2010}
J. S. Park, J. P. Kim, C. Song, M. Lee, J. S. Park, J. P. Kim, C. Song, M. Lee, Displays \textbf{31}, 164 (2010).

\bibitem{Deegan1997}
R. D. Deegan, O. Bakajin, T. F. Dupont, G. Huber, S. R. Nagel, T. A. Witten, Nature \textbf{389}, 827 (1997).

\bibitem{Osthoff2007}
R. C. Osthoff, S.W. Kantor, \textit{Organosilazane Compounds} John Wiley \& Sons, Inc.; (1997)

\bibitem{HandchemLide}
D. R. Lide, In \textit{Handbook of Chemistry and physics 86th ed.}; CRC Press Inc.; Boca Raton, FL, (2005)

\bibitem{makprl08}
K. F. Mak, M. Y. Sfeir, Y. Wu, C. H. Lui, J. A. Misewich,T. F. Heinz, Phys. Rev. Lett. \textbf{101}, 196405 (2008).

\bibitem{kravetsprb10}
V. G. Kravets, A. N. Grigorenko, R. R. Nair, P. Blake, S. Anissimova, K. S. Novoselov, A. K. Geim, Phys. Rev. B \textbf{81}, 155413 (2010).

\bibitem{Nair2008}
R. R. Nair, P. Blake, A. N. Grigorenko, K. S. Novoselov, T. J. Booth, T. Stauber, N. M. R. Peres, A. K. Geim, Science \textbf{320}, 1308 (2008).

\bibitem{Casiraghi2007}
C. Casiraghi, A. Hartschuh, E. Lidorikis, H. Qian, H. Harutyunyan, T. Gokus, K. S. Novoselov, A. C. Ferrari, Nano Lett., \textbf{7}, 2711 (2007).

\bibitem{Meyer_nature2007}
J. C. Meyer, A. K. Geim, M. I. Katsnelson, K. S. Novoselov, T. J. Booth, S. Roth, Nature \textbf{446}, 60 (2007).

\bibitem{Meyer_SSC2007}
J. C. Meyer, A. K. Geim, M. I. Katsnelson, K. S. Novoselov, D. Obergfell, S. Roth, C. Girit, A. Zettl, Solid State Commun. \textbf{143}, 101 (2007).

\bibitem{Ferrari2006}
A. C. Ferrari, J. C. Meyer, V. Scardaci, C. Casiraghi, M. Lazzeri, F. Mauri, S. Piscanec, D. Jiang, K. S. Novoselov, S. Roth, A. K. Geim, Phys. Rev. Lett. \textbf{97}, 4 (2006).

\bibitem{Khan2010}
U. Khan, A. O'Neill, M. Lotya, S. De, J. N. Coleman, Small \textbf{6}, 864 (2010).

\bibitem{hansenbook}
C. M. Hansen, \textit{Hansen Solubility Parameters: A User's Handbook}, CRC Press Inc., Boca Raton, FL (2007)

\bibitem{Bergin2008}
S. D. Bergin, V. Nicolosi, P. V. Streich, S. Giordani, Z. Sun, A. H. Windle, P. Ryan, N. P. P. Niraj, Z.-T. Wang, L. Carpenter, W. J. Blau, J. J. Boland, J. P. Hamilton, J. N. Coleman  Adv. Mater. \textbf{20}, 1876 (2008).

\bibitem{Lotya2010}
M. Lotya, P. J. King, U. Khan, S. De, J. N. Coleman, ACS Nano \textbf{4}, 3155 (2010).

\bibitem{Williams1958}
J. W. Williams, K. E. Van Holde, R. L. Baldwin, H. Fujita, Chem. Rev. \textbf{58}, 715 (1958).

\bibitem{Schuck2000}
P. Schuck, Biophys. J. \textbf{78}, 1606 (2000).

\bibitem{Svedberg1940}
T. Svedberg, K. O. Pedersen, \textit{The Ultracentrifuge}, Oxford University press, London (1940)

\bibitem{Ferrari2000}
A. C. Ferrari, J. Robertson, Phys. Rev. B \textbf{61},  14095 (2000).

\bibitem{Tuinstra1970}
F. Tuinstra, J. L. Koenig, J. Chem. Phys. \textbf{53}, 1126 (1970).

\bibitem{Piscanec2004}
S. Piscanec, M. Lazzeri, F. Mauri, A. C. Ferrari, J. Robertson, Phys. Rev. Lett. \textbf{93}, 4 (2004).

\bibitem{Casiraghi2009}
C. Casiraghi, A. Hartschuh, H. Qian, S. Piscanec, C. Georgi, A. Fasoli, K. S. Novoselov, D. M. Basko, A. C. Ferrari, Nano Lett. \textbf{9}, 1433 (2009).

\bibitem{Ferrari2001}
A. C. Ferrari, J. Robertson, Phys. Rev. B \textbf{64}, 13 (2001).

\bibitem{Cancado2011}
L.G. Cancado, A. Jorio, E. H. Ferreira, F. Stavale, C. A. Achete, R. B. Capaz, M. V. O. Moutinho, A. Lombardo, T. S. Kulmala, A.C. Ferrari,  Nano Lett., \textbf{11}, 3190 (2011).

\bibitem{Ferrari2003}
A. C. Ferrari, S. E. Rodil, J. Robertson, Phys. Rev. B \textbf{67}, 155306 (2003)

\bibitem{Ferrari2004}
Ferrari A. C., Surf. Coat. Tech. \textbf{180-181}, 190 (2004).

\bibitem{Basko2009}
D. M. Basko, S. Piscanec, A. C. Ferrari, Phys. Rev. B \textbf{80}, 165413 (2009).

\bibitem{Das2008}
A. Das, S. Pisana, B. Chakraborty, S. Piscanec, S. K. Saha, U. V. Waghmare, K. S. Novoselov, H. R. Krishnamurthy, A. K. Geim, A. C. Ferrari, A. K. Sood, Nat. Nano. \textbf{3}, 210 (2008).

\bibitem{Pisana2007}
S. Pisana, M. Lazzeri, C. Casiraghi, K. S. Novoselov, A. K. Geim, A. C. Ferrari, F. Mauri, Nat. Mater. \textbf{6}, 198 (2007).

\bibitem{Kaye1997}
B. H. Kaye, \textit{Powder mixing}; Chapman \& Hall; London, (1997).

\bibitem{Kauffman2001}
G. W. Kauffman, P. C. Jurs, J. Chem. Inf. Comp. Sci. \textbf{41}, 408 (2001).

\bibitem{Epson2011}
http://www.epson.com/cgi-bin/Store/Landing/ InkTechCartridges.jsp

\bibitem{Young1805}
T. Young, Philos. T. R. Soc. Lon. \textbf{95}, 65 (1805).

\bibitem{Shafrin1967}
E. G. Shafrin, W. A. Zisman,  J. Phys. Chem. \textbf{71}, 1309 (1967).

\bibitem{Thomas1996}
R. R. Thomas, F. B. Kaufman, J. T. Kirleis, R. A. Belsky, J. Electrochem. Soc., \textbf{143}, 643 (1996).

\bibitem{Glendinning1991}
W. B. Glendinning, J. N.  Helbert, \textit{Handbook of VLSI microlithography: principles, technology, and applications}, Noyes, New Jersey, (1991).

\bibitem{Ghatee2005}
M. H. Ghatee, L. Pakdel, Fluid Phase Equilibr. \textbf{234}, 101 (2005).

\bibitem{Marmur2003}
A. Marmur, Langmuir \textbf{19}, 8343 (2003).

\bibitem{Duineveld2003}
P. C. Duineveld, J. Fluid Mech. \textbf{477}, 175 (2003).

\bibitem{Gamerith2007}
S. Gamerith, A. Klug, H. Scheiber, U. Scherf, E. Moderegger, E. J. W. List, Adv. Func. Mater. \textbf{17}, 3111 (2007).

\bibitem{Smits1958}
F. M. Smits, Bell Sys. Tech. Jour. \textbf{37}, 711 (1958)

\bibitem{Hu2004}
L. Hu, D. S. Hecht, G. Gruner , Nano Lett. \textbf{4}, 2513 (2004).

\bibitem{Geng2007}
H. Z. Geng, K. K. Kim, K. P. So, Y. S. Lee, Y. Chang, Y. H. Lee  J. Am. Chem. Soc. \textbf{129}, 7758 (2007).

\bibitem{DeACS2010}
S. De, P.J. King, P.E. Lyons, U. Khan, J. N. Coleman, ACS Nano \textbf{4} 7064 (2010).

\bibitem{DeSmall2009}
S. De, J. N. Coleman, Small \textbf{6}, 458 (2009).

\bibitem{Kirkpatrick1973}
S. Kirkpatrick, Rev. Mod. Phys. \textbf{45}, 574 (1973).

\bibitem{Stauffer1985}
D. Stauffer, A. Aharony, \textit{Introduction to percolation theory}, Taylor\&Francis: London, (1985).

\bibitem{Kogut1979}
P. M. Kogut, J. P. Straley, J. Phys. C \textbf{12}, 2151 (1979).

\bibitem{Johner2008}
N. Johner, C. Grimaldi, I. Balberg, P. Ryser, Phys. Rev. B \textbf{77}, 174204 (2008)

\bibitem{Doherty2009}
E. M. Doherty, S. De, P. E. Lyons, A. Shmeliov, P. N. Nirmalraj, V. Scardaci, J. Joimel, W. J. Blau, J. J. Boland, J. N. Coleman, Carbon \textbf{47}, 2466 (2009)

\bibitem{Kuzmenko2008}
A. B. Kuzmenko, E. Van Heumen, F. Carbone, D. van der Marel, Phys. Rev. Lett. \textbf{100}, 117401 (2008).

\bibitem{Ong2004}
B. S. Ong, Y. Wu, P. Liu, S. Gardner, J. Am. Chem. Soc. \textbf{126}, 3378 (2004)

\bibitem{Arias2004}
A. C. Arias, S. E. Ready, R. Lujan, W. S. Wong, K. E. Paul, A. Salleo, M. L. Chabinyc, R. Apte, R. A. Street, Y. Wu, P. Liu, B. Ong, Appl. Phys. Lett. \textbf{85}, 3304 (2004).

\bibitem{Wu2005}
Y. Wu, P. Liu, B. S. Ong, T. Srikumar, N. Zhao, G. Botton, S. Zhu, Appl. Phys. Lett. \textbf{86}, 142102 (2005)

\bibitem{Kaneto2007}
K. Kaneto, M. Yano, M. Shibao, T. Morita, W. Takashima, Jap J. App. Phys. \textbf{46}, 1736 (2007)

\bibitem{Morita2010}
T. Morita, V. Singh, S. Oku, S. Nagamatsu, W. Takashima, S. Hayase, K. Kaneto, Jap. J. App. Phys. \textbf{49}, 04161 (2010)

\bibitem{Oh2009}
J. H. Oh, H. W. Lee, S. Mannsfeld, R. M. Stoltenberg, E. Jung, Y. W. Jin, J. M. Kim, J.-B. Yoo, Z. Bao,  PNAS \textbf{106}, 6065 (2009).

\bibitem{Sirringhaus1998}
H. Sirringhaus, N. Tessler, R. H. Friend, Science \textbf{280}, 1741 (1998)

\bibitem{Song2010}
Y. J. Song, J. U. Lee, W. H. Jo, J. Song, J. U. Lee, W. H. Jo, Carbon \textbf{48}, 389 (2010)

\bibitem{Whiting2009}%
G. L. Whiting, A.C. Arias,  Appl. Phys. Lett. \textbf{95}, 253302 (2009)

\bibitem{Klauk_book2006}
H. Klauk, \textit{Organic Electronics}, Wiley-VCH: Weinheim, (2006); Chapter 4.

\bibitem{Chason2005}%
M. Chason, P.W. Brazis, J. Zhang, K. Kalyanasundaram, D. R. Gamota,  Proc. IEEE \textbf{93}, 1348 (2005)

\bibitem{Kawase2005}
T. Kawase, S. Moriya, C. J. Newsome, T. Shimoda,  Jap. J. App. Phys. \textbf{44}, 3649 (2005)

\bibitem{Parmer2008}
J. E. Parmer, A. C. Mayer, B. E. Hardin, S. R. Scully, M. D. McGehee, M. Heeney, I. McCulloch,  Appl. Phys. Lett. \textbf{92}, 113309 (2008)

\bibitem{Padday1971}
J. F. Padday, Phyl. Trans. R. Soc. Lond. A \textbf{269}, 265 (1972)

\end{thebibliography}
\end{document}